\documentclass[prl,twocolumn,showpacs,preprintnumbers,amsmath,amssymb,floatfix]{revtex4}
\usepackage{graphicx}
\usepackage{dcolumn}
\usepackage{bm}

\usepackage{times}
\usepackage[T1]{fontenc}
\usepackage[latin1]{inputenc}

\begin{document}
\title{Self-Assembly of Information in Networks}

\author{M. Rosvall$^{1,2}$}\email{rosvall@tp.umu.se}
\author{K. Sneppen$^{2}$}

\affiliation{$^{1)}$Department of Theoretical Physics, Ume{\aa} University,
901 87 Ume{\aa}, Sweden\\
$^{2)}$Niels Bohr Institute, Blegdamsvej 17, Dk 2100, Copenhagen, Denmark}
\homepage{http://cmol.nbi.dk}

\date{\today}

\begin{abstract}
We model self-assembly of information in networks to investigate 
necessary conditions for building a global perception of a system by local communication.
Our approach is to let agents chat in a model system
to self-organize distant communication-pathways.
We demonstrate that simple local rules allow agents
to build a perception of the system, that is robust
to dynamical changes and mistakes.
We find that messages are most effectively
forwarded in the presence of hubs, while transmission in hub-free networks
is more robust against misinformation and failures.
\end{abstract}

\maketitle
Communication is essential in systems
ranging from human society to mobile telephone- and computer networks.
It enables parts of a system to build a global perception,
and thereby makes it possible for these parts to overcome the
information horizon\cite{friedkin,valverde1,trusina2004}
set by their immediate neighbors.
We mimic real-world situations to investigate what limits
the local generation of this global perception of the
network from multiple communication events.
Our approach is to let agents chat in a model system
to self-organize distant communication-pathways,
and thereby make use of the typical small-world properties of networks\cite{watts}.
We investigate the necessary conditions for building a global perception,
and demonstrate that simple local rules allow agents
to build a perception of the system that is robust
to dynamical changes and mistakes.
In this minimalistic model, we find that messages are most effectively
forwarded in the presence of hubs with funnelling\cite{milgram1969},
like in scale-free networks, while transmission in hub-free networks
is more robust against misinformation and failures.

\begin{figure}
\begin{center}
\leavevmode
\includegraphics[width=1.0\columnwidth]{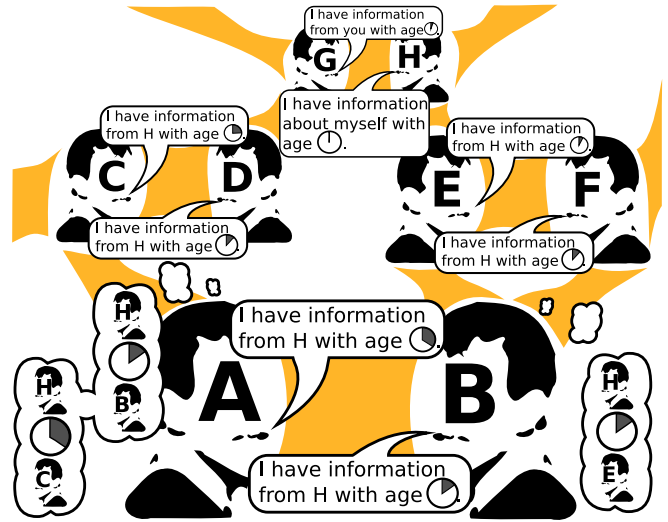}
\end{center}
\caption{\label{fig1}(Color online) Self-assembly of information as modeled in this
paper. Agents at nodes communicate with their acquaintances about any third target agent
in the network, and estimate the quality of the information by its age.
Here, agent \textbf{A} learns that \textbf{B} has newer information about
\textbf{H}, disregards its old information, and change its pointer associated to \textbf{H} from \textbf{C} to \textbf{B}.
The three memory bubbles from left to right, the information about \textbf{H} of \textbf{A} before and after the communication event, and the information about \textbf{H} of \textbf{B} represent:
The target agent (top), the age of the information (middle), and the acquaintance that provided the information (bottom).
Every agent has a corresponding perception for all other agents in the network.}
\end{figure}

To visualize our basic approach we illustrate in Fig.\ \ref{fig1}
the rules of communication in a network composed of individual agents, each
of them connected to a number of acquaintances. Each individual communicates 
with its immediate neighbors in order to exchange information about agents
in other parts of the system. In this way every individual gradually builds up a
global perception by knowing people through people\cite{friedkin-infoflow}.
This can be modeled in terms of agents having information about the position of 
other agents in the system. In our minimalistic model,
we allow each agent to have information 
about which of its neighbors that connects most
efficiently to each of the other agents in the system.
Thus, a perfectly informed agent will know which direction to send a 
message to any other agent in the system. If all agents were perfectly informed,
any message would be reliably forwarded from sender to recipient, using the
information of the subsequent agents along its paths.
Through these communications, agents create a simplistic routing protocol,
related to the pro-active protocols for mobile networks\cite{perkins}.
This is a routing procedure that, in its simplest form, only requires agents 
to know to whom to transfer information\cite{milgram1969,duncan}.
Our approach opens up for schematic modeling of self-assembly of
information in a variety of systems, including social systems.

\begin{figure}
\begin{center}
\leavevmode
\includegraphics[width=\columnwidth]{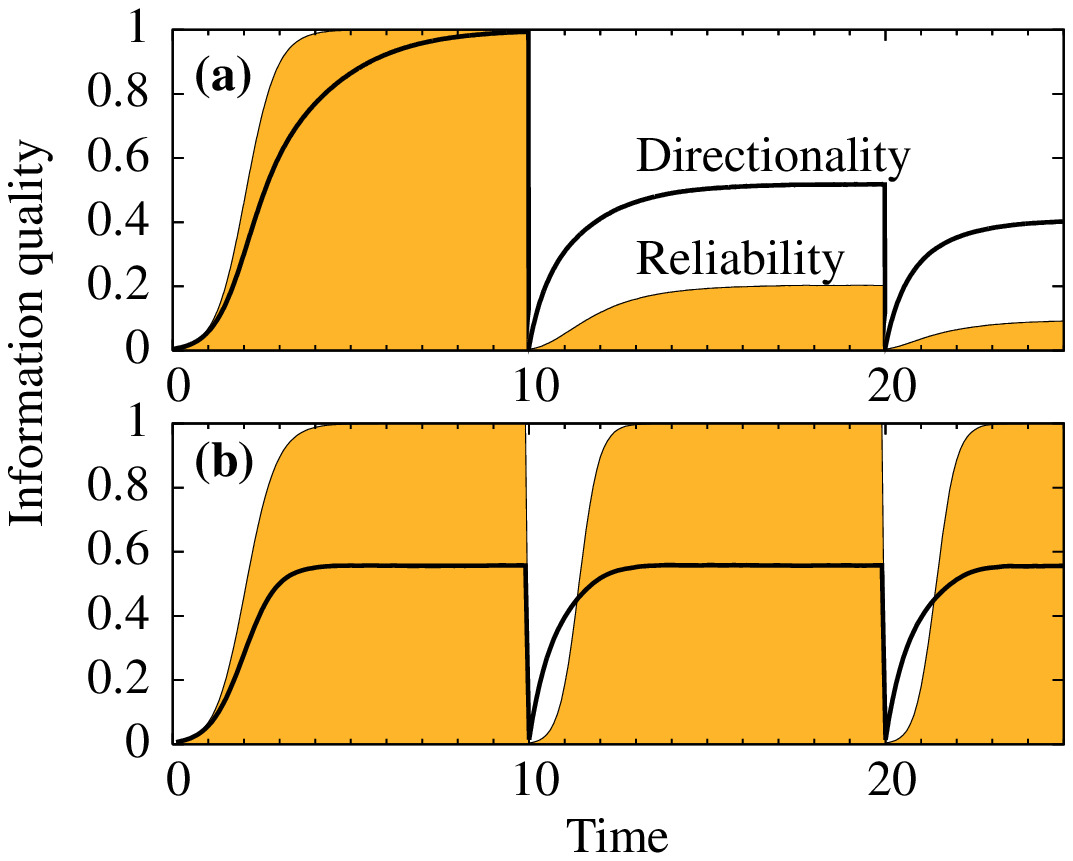}
\end{center}
\caption{\label{fig2}(Color online)
The quality of the self-assembled information.
The reliability (filled curve), is the fraction of messages
that reach their targets guided by the agents' pointers .
The directionality (black line), is the fraction of pointers that
points to an acquaintance that is closer to the target agent than the agent itself 
The time unit is $N$ communication events per agent.
The network perception develops from complete absence of information at time 0.
The quality of the self-assembled information is in (a) estimated by distance and 
in (b) by time.
At time 10 and time 20, all links are completely reshuffled.
The agents thereby get wrong perception,
but the reliability is recovered within a few time units in the time-based
update in (b), but not in the distance-based update in (a).
The simulation is performed on a connected Erd{\H o}s-Rényi network \cite{erdos},
with $N=1000$ nodes and $L=2000$ links.
}
\end{figure}

The key question is how different communication rules of the
agents influence their possibility to obtain a reliable perception
of the network. Obviously, if the two acquaintances \textbf{A} and \textbf{B} in Fig.\ \ref{fig1} just
exchange information about the possible directions to a target, say agent \textbf{H},
they may agree or disagree, but cannot decide which
information is best. As a result, the correct
information is not transmitted, and no coherent
perception can emerge. The agents need some index of quality that let
them judge who has the best knowledge.
One option is that each agent both have a pointer
(the acquaintance that provided the information)
and an estimate of the number of intermediates there are on
the information path to the target agent \textbf{H} \cite{rosvall}.
In social systems this can be motivated by the observation
that knowledge of a person is related to the shortest path\cite{friedkin}.
For example, if two acquaintances communicate about a third agent, the 
agent with the longer distance estimate simply adopts the view of its acquaintance.
The agent change the information about the target agent,
sets its pointer to the acquaintance and its distance estimate to the acquaintance's estimate plus 1.
With this method all agents obtain perfect knowledge of both directions and distance
in any type of statically connected network, see Fig.\ \ref{fig2}(a).

The use of distances as quality of information about directions has
some important costs. In particular, we found that if someone somewhere gives wrong
information and provides a distance that is smaller than the real
one, this misinformation will overrule the correct information and lead to
permanent damages. The method is exact to the price of
not being robust to any structural changes (or other sources
of wrong information), see Fig.\ \ref{fig2}(a).

We therefore introduce another way of validating information
to mimic systems that should be robust to dynamical changes:
The age of the information about a target agent,
say agent \textbf{H} in Fig.\ \ref{fig1}.
When an acquaintance of \textbf{H} obtains information about \textbf{H}, it
sets its pointer to \textbf{H}, and the information starts aging.
With successive communication events,
the information spreads from agent to agent and gets older and older
(we increase the age of all information when all links on average have participated in one communication event).
When two agents compare the validity of their pointers to a target agent,
like \textbf{A} and \textbf{B} to \textbf{H} in Fig.\ \ref{fig1},
they validate the newest information as the most correct one.

Figure \ref{fig2}(b) shows the quality of the self-assembled information.
The result was obtained by putting 1000 agents without any prior information 
on a connected Erd{\H o}s-Rényi network \cite{erdos},
followed by successive communication events:
A randomly chosen link connects two agents that
communicate about a third randomly chosen target agent.
The communication rules make the agents point in the direction
of the fastest communication path from a target,
and we only use the shortest path as a benchmark.
The directionality is the fraction of all pointers 
for all agents that points to an acquaintance that
is closer to the target agent than the agent itself.
The directionality never reaches 100\%,
because some paths between two agents are updated faster than the shortest path.
The message paths depend on the communication activity and fluctuate with time,
but anyway stay fairly close to the shortest paths (see Fig.\ \ref{fig4}(b) and Java-simulation \cite{java}).
To investigate the reliability of such forwarded messages,
we selected all possible combinations of pairs of agents as sources and targets,
and released messages at the sources.
The reliability in Fig.\ \ref{fig2}(b) shows that 100\% of the
messages reached their targets guided by the agents' pointers along the paths (when the information was recovered).
The main feature of this time-based update is that the agents forget old mistakes,
and that their network perception recovers completely after reshuffling of the links (see Fig.\ \ref{fig2}(b)).
We find that the reliability is insensitive to network topology or details in the model,
but that the recovery and the communication paths depend on communication habits.

\begin{figure}
\begin{center}
\leavevmode
\includegraphics[width=1.0\columnwidth]{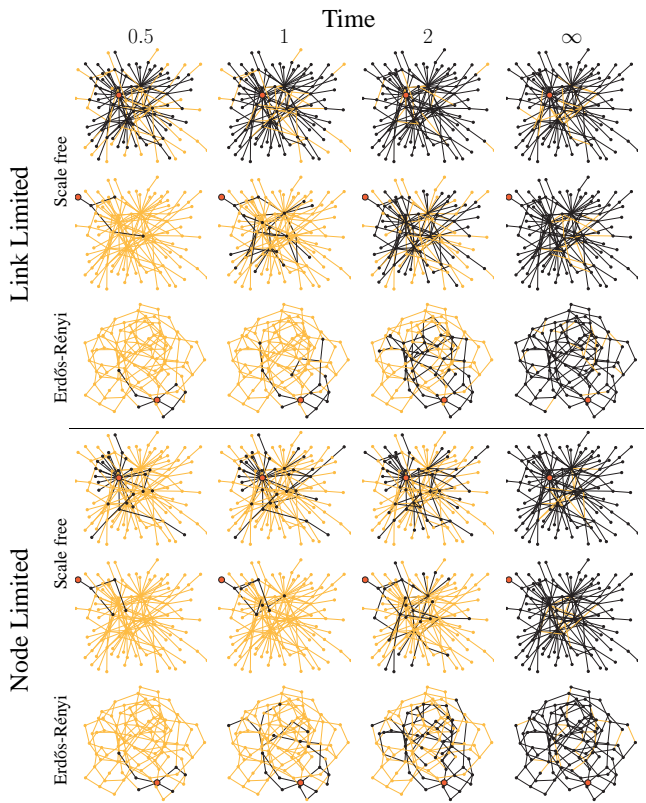}
\end{center}
\caption{\label{fig3}(Color online)
Initial spread of information.
The black parts of the network show paths which are
known by the red-marked agent. As time progresses the red agent builds its perception
until complete coverage of all nodes (not all links) in the network.
The upper panel investigates a setup where each agent communicates with a
frequency proportional to its connectivity (LL), whereas the lower panel
shows the more restricted case where each agent communicate equally
much (NL). To explore the effects of well connected agents,
we compare scale-free with Erd{\H o}s-Rényi networks
($N=100$ nodes and $L=160$ links),
and hub nodes with peripheral nodes.
An interactive Java simulation is available online \cite{java}.
}
\end{figure}

In implementing the communication model we also
specify how often different nodes are selected for communication.
Above we did so by simply assuming that persons with many
connections were more active in communication (by choosing a random link).
This will be referred to as link limited communication, LL, in contrast
to the case where each node is equally active, denoted
node limited communication, NL.
In Fig.\ \ref{fig3}, we illustrate how individual nodes
use communication to build up a perception of the
surrounding network (this is visualized by letting the agents transfer
the full path to the target node at each communication event, see also simulation at \cite{java}).
This can also represent the spread of information about individual nodes
in initially non-informed networks. We use a scale-free (SF) network as
an extreme, yet realistic \cite{faloutsos,barabasi},
example of networks with hubs, and an Erd{\H o}s-Rényi (ER)
network as an example of networks without hubs\cite{erdos}.
The upper two sequences in Fig.\ \ref{fig3} illustrate the information
spreading around, respectively, a hub node, and a single-linked node.
We see that hubs most rapidly gain information about a
large part of the network \cite{friedkin-infoflow}.
The third sequence illustrates information
spreading in a network without hubs.
The dynamic advantage of scale-free networks, with faster self-assembly of information,
is closely associated to the higher communication activity of hubs (funneling, see \cite{milgram1969}).
In the node limited version, where each agent
communicates equally much, the advantages of hubs are severely reduced.

\begin{figure}
\begin{center}
\leavevmode
\includegraphics[width=1.0\columnwidth]{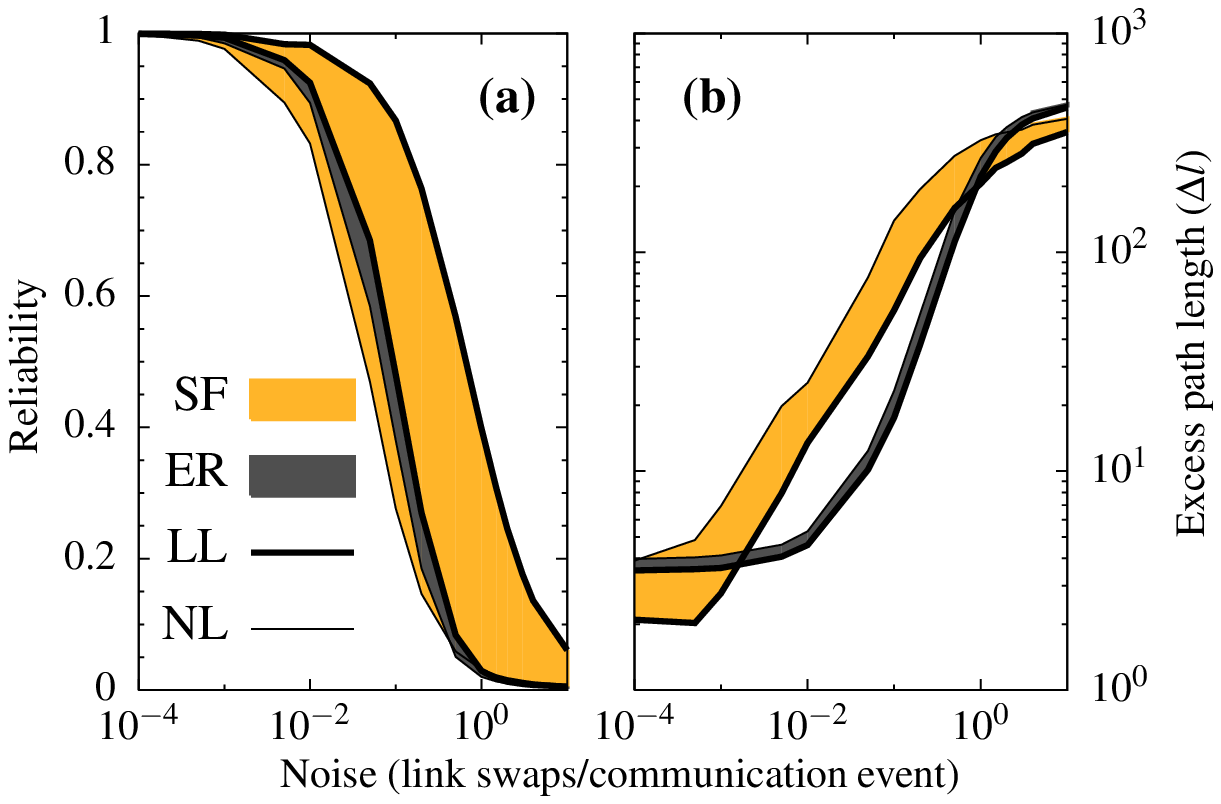}
\end{center}
\caption{\label{fig4}(Color online)
The ability to self assemble information in a noisy environment.
The unit of noise is one link swap\cite{maslov2002} per communication event.
As in Fig.\ \ref{fig2} we in (a) show the reliability of forwarded messages,
here for a scale-free (SF) and an Erd{\H o}s-Rényi (ER) network with communication
limited to, respectively, links (LL) and nodes (NL). The network size is $N=1000$ nodes
and $L=2000$ links and the scale-free network is generated with the degree distribution
$P(k) \propto k^{-2.4}$ with the method suggested in\cite{trusina-hierarchy}.
$\Delta l/N$ in (b) is the average excess path length of forwarded 
messages compared to the shortest path length between agents.}
\end{figure}

We now turn to how messages are transmitted in the model networks
with varying degree of false information (Fig.\ \ref{fig4}).
We parametrize noise in terms of dynamically changing networks,
where pairs of links are swapped \cite{maslov2002}
at different ``boiling'' rates.
A directed message might get trapped in a closed loop,
because the pointers are not always updated.
Figure \ref{fig4}(a) shows the reliability of
forwarded messages in networks with hubs (SF) and without hubs (ER) for communication
limited to, respectively, links (LL) and nodes (NL).
We see that the communication habits do
not affect the performance on the hub-free Erd{\H o}s-Rényi network.
With, and only with the large communication ability of hubs (LL),
scale-free networks provide the most reliable communication.

In Fig.\ \ref{fig4}(b), we go one step further and investigate 
the excess path length $\Delta l$ of messages,
by comparing with the shortest-path length between agents.
Again we choose all possible combinations of source and target agents.
To avoid being trapped in closed loops, we here let the messages step 
to a random acquaintance when it reaches an agent the second time
between a source and a target agent.
Obviously $\Delta l$ increases with increasing noise level.
However, and more importantly, at a wide range of boiling
rates, $\Delta l$ for scale-free networks is substantially larger
than $\Delta l$ for networks without hubs.
Scale-free networks are less robust signal transmitters than networks without hubs.
Hubs are sometimes efficient, but they tend to accumulate mistakes,
and are not necessary to provide short paths \cite{dodds}.
Contrary, networks without hubs provide many alternative short paths.

The social implications of our model may be illustrated
through an analogy to Milgram's famous experiment where letters
were transmitted by sequences of person-person contacts across USA\cite{milgram,milgram1969}.
The surprising result was that
the letters, that actually arrived, only used a few intermediate
contacts. Since the dimension in social networks is high\cite{kochen},
the non-trivial result was not that short paths existed,
but that they were found in the experiment.
The accepted explanation is, again that world is small\cite{kochen,watts},
and that people use geographic closeness of the acquaintance to the target (first steps)
and similarity of occupation (later steps) to forward messages\cite{killworth,dodds,duncan,kleinberg-hierarchy}.
This is in overall accordance with our scenario
where messages that arrive typically use short paths,
and where any serious deviation leads to much
longer paths and in practice to lost messages.
Obviously, one could add a number of
layers to our minimalistic model \cite{java}, including in particular:
Geographic closeness, social identity, and interest spheres\cite{killworth,dodds,duncan,kleinberg-hierarchy},
strength of links \cite{granovetter},
and information decay \cite{waugh} or dropouts at the passage of messages\cite{milgram,hunter}.
In any case, we found that our simple interaction rules give a perception that is enough to
overcome the information horizon\cite{friedkin} set by immediate acquaintances.
It is a framework for further exploration and a possible bridge between network theoretical
approaches \cite{watts,kleinberg} and experiments \cite{milgram,hunter,dodds}.

Social as well as many modern information networks are highly complex.
In addition to the dynamical characteristics of both the local properties and the global organization,
they also have a complex interplay.
Nevertheless, the systems should have the ability to self-organize their locally available information
such that messages can be guided between distant parts of the network.
To study such self organization, we have presented
a generic framework that also is open for a direct extension to networks where
the capacity of links or nodes vary with time.
When some pathways are temporarily occluded by congestion\cite{huberman} or malfunction\cite{barabasi-attack},
the local chatting allows the system to adapt dynamically to
the new situation and subsequently transmit messages along other fast pathways.


\bibliographystyle{Science}

\end{document}